\documentclass[iop,revtex4]{emulateapj}
\usepackage{amsmath}
\usepackage{ulem}
\usepackage[dvips, usenames]{color}

\shorttitle{Near-infrared DIBs toward Cyg OB2}
\shortauthors{Hamano et al.}

\begin{document}

\title{Near infrared diffuse interstellar bands toward the Cygnus OB2 association}
\author{Satoshi Hamano\altaffilmark{1},Naoto Kobayashi\altaffilmark{1,2,3}, Sohei Kondo\altaffilmark{1}, Hiroaki Sameshima\altaffilmark{1}, Kenshi Nakanishi\altaffilmark{1}, Yuji Ikeda\altaffilmark{1,4}, Chikako Yasui\altaffilmark{1,5}, Misaki Mizumoto\altaffilmark{5,6}, Noriyuki Matsunaga\altaffilmark{1,5}, Kei Fukue\altaffilmark{5}, Ryo Yamamoto\altaffilmark{5}, Natsuko Izumi\altaffilmark{5}, Hiroyuki Mito\altaffilmark{1,3}, Tetsuya Nakaoka\altaffilmark{1,7}, Takafumi Kawanishi\altaffilmark{1,7}, Ayaka Kitano\altaffilmark{1,7}, Shogo Otsubo\altaffilmark{1,7}, Masaomi Kinoshita\altaffilmark{8} and Hideyo Kawakita\altaffilmark{1,7}}
\email{hamano@cc.kyoto-su.ac.jp}

\altaffiltext{1}{Laboratory of Infrared High-resolution spectroscopy (LIH), Koyama Astronomical Observatory, Kyoto Sangyo University, Motoyama, Kamigamo, Kita-ku, Kyoto 603-8555, Japan}
\altaffiltext{2}{Institute of Astronomy, School of Science, University of Tokyo, 2-21-1 Osawa, Mitaka, Tokyo 181-0015, Japan}
\altaffiltext{3}{Kiso Observatory, Institute of Astronomy, School of Science, The University of Tokyo, 10762-30 Mitake, Kiso-machi, Kiso-gun, Nagano, 397-0101, Japan}
\altaffiltext{4}{Photocoding, 460-102 Iwakura-Nakamachi, Sakyo-ku, Kyoto, 606-0025, Japan}
\altaffiltext{5}{Department of Astronomy, Graduate School of Science, University of Tokyo, Bunkyo-ku, Tokyo 113-0033, Japan}
\altaffiltext{6}{Institute of Space and Astronautical Science (ISAS), Japan Aerospace Exploration Agency (JAXA), 3-1-1, Yoshinodai, Chuo-ku, Sagamihara, Kanagawa, 252-5210, Japan}
\altaffiltext{7}{Department of Physics, Faculty of Sciences, Kyoto Sangyo University, Motoyama, Kamigamo, Kita-ku, Kyoto 603-8555, Japan}
\altaffiltext{8}{Solar-Terrestrial Environment Laboratory, Nagoya University, Furo-cho, Chikusa-ku, Nagoya, Aichi, 464-8601, Japan}
\begin{abstract}

We obtained the near-infrared (NIR) high-resolution ($R\equiv\lambda/\Delta\lambda\sim20,000$) spectra of the seven brightest early-type stars in the Cygnus OB2 association for investigating the environmental dependence of diffuse interstellar bands (DIBs). The WINERED spectrograph mounted on the Araki 1.3m telescope in Japan was used to collect data. All 20 of the known DIBs within the wavelength coverage of WINERED ($0.91<\lambda<1.36\mu$m) were clearly detected along all lines of sight because of their high flux density in the NIR wavelength range and the large extinction. The equivalent widths (EWs) of DIBs were not correlated with the column densities of C$_2$ molecules, which trace the patchy dense component, suggesting that the NIR DIB carriers are distributed mainly in the diffuse component. On the basis of the correlations among the NIR DIBs both for stars in Cyg OB2 and stars observed previously, $\lambda\lambda$10780, 10792, 11797, 12623, and 13175 are found to \textcolor{black}{constitute} a ``family'', in which the DIBs are correlated well over the wide EW range. In contrast, the \textcolor{black}{EW} of $\lambda$10504 is found to remain almost constant over the stars in Cyg OB2. The extinction estimated from the average EW of $\lambda$10504 ($A_V\sim3.6$mag) roughly corresponds to the lower limit of the extinction distribution of OB stars in Cyg OB2. This suggests that  $\lambda$10504 is absorbed only by the foreground clouds, implying that the carrier of $\lambda$10504 is completely destroyed in Cyg OB2, probably by the strong UV radiation field. The different behaviors of the DIBs may be caused by different properties of the DIB carriers. 

\end{abstract}


\section{Introduction}

Diffuse interstellar bands (DIBs) are ubiquitous absorption \textcolor{black}{bands} detected in the spectra of reddened stars. 
Recent surveys have discovered more than 500 DIBs in the near-UV to near-infrared (NIR) wavelength range \citep{jen94b,hob08,hob09,geb11,ham15}. Although no DIB carriers have been successfully identified, the absorption features might mainly be contributed by organic compounds \citep{sar08}. Recently, some reseachers have attempted to directly identify the DIB carriers by measuring the absorption spectra of gas-phase organic molecules in the laboratory \citep{mai11,sal11,gre11}. \citet{cam15}, in the first study regarding the successful identification of DIB carriers, confirmed that ionized buckminsterfullerene (C$_{60}^+$) is responsible for two DIBs at $\lambda=$9577 and 9633 \r{A}. These bands were originally proposed to coincide with the spectral features of C$_{60}^+$ \textcolor{black}{by \citet{foi94}}. 

Elucidating the behavior of DIBs in various environments would help us to identify the properties of the carriers \citep[e.g.,][]{jen94a,vos11}, and possibly enable new diagnostics of the physical parameters of gas clouds. The DIB equivalent width (EW) ratio $\lambda 5780/ \lambda5797$ is known to depend on the UV radiation field \citep{cam97,vos11,kos13}. Because these DIBs are easily observed by optical spectroscopy, this ratio can indicate the UV radiation field of the gas clouds toward, for example, the extragalactic lines of sight \citep{sol05}. In dense clouds, some strong DIBs (e.g., $\lambda \lambda 5780, 5797$ and 6614) are relatively weaker than what is expected from the $A_V$ correlations. This is considered to be because the molecules are shielded from UV photons \citep{ada91,ada94,sno02,ada05}. On the other hand, some weak DIBs are suggested to be well correlated with C$_2$ column densities \citep{tho03,ada05}. \citet{wel06} investigated DIBs originating from the interstellar clouds in the Magellanic Clouds, reporting that the three strong DIBs $\lambda \lambda 5780, 5797$, and 6284 become weaker by factors of $\sim$ \textcolor{black}{7--9} and $\sim 20$, respectively, toward the Large and Small Magellanic Clouds than in the Galactic interstellar environment \citep[see also][]{ehr02,sol05,cox06}. They suggested that the DIBs are weakened by the lower metallicity and higher UV radiation in the Magellanic Clouds. \textcolor{black}{[1] The relations between DIBs and the other interstellar features, such as \ion{Na}{1} D lines, \ion{Ca}{2} H, K lines, and dust emission, have been examined by comparing their spatial distributions: \citet{van09} for extra-planer gas, \citet{far15} and \citet{bai15b} for the Local Bubble, and \citet{van13} and \citet{bai15a} for LMC. As such, these extensive new observations are gradually shedding new light on the dependence of DIBs on environment. }

NIR observations are advantageous for exploring DIBs toward stars with high extinctions. Therefore, they should also prove useful for investigating the environment-dependent properties of DIBs \citep{ada94}. NIR DIB observations have been limited because the performance of NIR high-resolution spectrographs (e.g., spectral resolution and sensitivity) is lower than that of optical spectrographs. However, NIR spectrography has progressed sufficiently to enable high-sensitivity searching for NIR DIBs \citep{geb11,cox14,raw14,zas15,ham15}. We are advancing the first comprehensive DIB survey in the NIR wavelength range using a newly developed high-resolution NIR spectrograph WINERED \citep{kon15}. In our first paper \citep[][hereafter Paper I]{ham15}, we successfully identified 15 new DIBs in the range $0.91 < \lambda < 1.32 \mu$m. Seven of these DIBs were reported as DIB candidates by \citet{cox14}. We suggested two DIB families (one consisting of $\lambda \lambda$10780, 10792, and 11797, and the other consisting of  $\lambda \lambda$10504, 12623, and 13175), whose EWs are highly correlated (\textcolor{black}{correlation coefficients} are $>0.9$; see Paper I). 

In this paper, we investigate the environmental dependence of NIR DIBs from the high-resolution spectra of seven early-type stars in Cyg OB2 newly obtained \textcolor{black}{with} the WINERED spectrograph. Cyg OB2, located at the distance of $\sim$1.5 kpc from the Sun \citep{han03}, is one of the most massive clusters or OB associations in our Galaxy. Statistical analysis of 2MASS survey data suggests as many as $\sim$ 2600 OB stars in this association \citep{kno00}. Despite its proximity, Cyg OB2 is known to be veiled with heavy interstellar extinction. This large extinction is partly ascribed to the dust lane called the Great Cygnus Rift, located at an estimated distance of 850 $\pm$ 25 pc from the Sun \citep[based on color-magnitude diagrams of the Cyg OB2 region; ][]{gua12}. According to \citet{wri15}, who calculated the individual extinctions of 164 OB stars (including our targets), the extinctions of Cyg OB2 members generally have a range of 4--7 mags, with a median $A_V$ of 5.4 mag, although the extinction of No.\,12 is anomalously large ($A_V = 10.2$ mag). The large spread of the extinctions might be caused by foreground extinction and/or intra-cluster extinctions \citep{wri15}. Cyg OB2 is known as one of the best lines of \textcolor{black}{[2] sight for detecting weak DIBs with high accuracy} because of its large extinction \citep{chl86}.

The gas clouds along lines of sight have been extensively investigated from the absorption lines of various small molecules, such as C$_2$, CN, CO, and H$_3^+$ \citep{mcc98,geb99,gre01}. The mm-wave emission lines of CO and CS have also been investigated \citep{sca02,cas05,sca07}. Because of its strong association with H$_3^+$ molecules, Cyg OB2 No.\,12 has attracted particular attention \citep{mcc98}. \citet{cec00} developed a model that reproduced the abundances of H$_3^+$, C$_2$ and CO toward No.\,12. According to this model, the H$_3^+$ absorption originates from the diffuse component, whereas C$_2$ and CO are formed in the embedded denser component. From the C$_2$ Phillips bands of No.\,12, \citet{gre01} derived the density as $n=600\pm 100$ cm$^{-3}$. Dense gas clouds are also supported by the observed CO emission lines \citep{sca02}. The abundances of CO, C$_2$, CN and CH can be explained by the chemistry induced under high X-ray fluxes from early-type stars in the association \citep{gre01}. As such, the clouds toward Cyg OB2 have complex density structures and are probably exposed to the strong fluxes of high-energy photons from OB stars, which would affect the EWs of the DIBs. 

The rest of this paper is organized as follows. Section \textsection{2} describes our observations, targets, and the data reduction procedures, and \textsection{3} describes our search for NIR DIBs. In \textsection{4} and \textsection{5}, we discuss the correlations among the NIR DIBs and  the environmental dependence of the NIR DIBs toward Cyg OB2, respectively. The study is summarized in \textsection{6}. 

\section{Observation, Targets, and Data Reduction}

Data were collected withs the newly developed high-resolution NIR echelle spectrograph, WINERED \citep{ike06,yas06,yas08,kon15}, mounted on the F/10 Nasmyth focus of the Araki 1.3 m telescope at Koyama Astronomical Observatory, Kyoto-Sangyo University, Japan \citep{yos12}. WINERED uses a 1.7 $\mu$m cutoff 2048 $\times$ 2048 HAWAII-2RG infrared array with a pixel scale of $0.''8$ pixel$^{-1}$, simultaneously covering the wavelength range 0.91--1.36 $\mu$m. We used a slit of length $48''$ and width $1.''6$ (2 pix). This slit width corresponds to a \textcolor{black}{spectral resolving power of $R\equiv \lambda / \Delta \lambda =28,000$} or $\Delta v = 11$ km s$^{-1}$. Immediately prior to observation in August 2014, we installed an $H$-band blocking filter to decrease the thermal leak in the 1.7--1.8 $\mu$m range. The new filter successfully decreased the background noise, yielding spectra with high signal-to-noise ratio (S/N). However, the new filter was slightly bent by the tight mechanical mounting, causing a slight off-focus of the light on the infrared array. Consequently, the spectral resolution became $R\sim20,000$, which is lower than the nominal value\footnote{This problem \textcolor{black}{is} now solved with an overhaul of the instrument.}. Because we here focus on the EWs of the DIBs, the lower spectral resolution does not significantly affect our results and following discussion.

We observed the seven brightest stars in the $J$-band in Cyg OB2 (No.\,3, 5, 8A, 9, 10, 11, and 12). The observations, which were made during September and October of 2014, are summarized in Table \ref{Cygtargets}. We also observed the telluric standard A0V-type stars at similar airmasses to the target airmasses. Although target No.\,12 was previously observed by us (Paper I), the engineering array in that observation introduced large systematic noise. In the present study we re-investigate No.\,12 for more accurate DIB EW estimates. Only the newly obtained data are analyzed in the present paper.

All data were reduced with IRAF using standard procedures, including sky subtraction, flat fielding, extraction of spectra, and wavelength calibration with Argon lamp spectra. To remove the telluric absorption lines, the target spectra were divided by those of the corresponding standard stars. Note that we did not remove the hydrogen and metal absorption lines in the standard stars, because these features are often contaminated by atmospheric absorption lines, which complicates their profile fitting and removal. Therefore, emission-like features appear in the divided spectra, which are fully identified and carefully considered when quantifying the EWs of the DIBs. For each echelle order, the spectra were normalized by a low-order Legendre function using the IRAF task, \textit{continuum}.

\begin{deluxetable*}{cccccccc}
\tabletypesize{\scriptsize}
 \tablecaption{Summary of Observations and Targets in Cyg OB2}
\tablewidth{0pt}
\tablehead{
\colhead{Stars} & \colhead{No.\,3} & \colhead{No.\,5} & \colhead{No.\,8A} & \colhead{No.\,9} & \colhead{No.\,10} & \colhead{No.\,11} & \colhead{No.\,12}
}
\startdata
Spectral Type & O9  &  O7Ia  &  O6If+O5.5III &  O5If  &  O9.5Ia  &  O5.5Ifc  &  B3-4Ia \\
$J$ (mag)& 6.498  &  5.187  &  6.123  &  6.468  &  6.294  &  6.650  &  4.667  \\
$E(B-V)$\footnote{The color excesses calculated from \citet{tor91}.} (mag) & 2.05  &  2.11  &  1.59  &  2.24  &  1.89  &  1.78  &  3.36  \\
Obs. Date & 2014/9/11  &  2014/9/20  &  2014/9/16  &  2014/9/13  &  2014/9/11  &  2014/9/14  &  2014/10/17  \\
Integration Time (sec) & 1,800  &  1,200   &  2,400  &  1,800   &  1,800  &  2,400   &  3.600   \\
S/N\footnote{The signal-to-noise ratio of the spectrum at $\lambda \sim 10400$\r{A}.} & 300  &  400  &  300  &  300  &  250  &  300  &  500 \\
S/N\footnote{The signal-to-noise ratio of the spectrum at $\lambda \sim 10400$\r{A} after the correction of telluric absorption lines.} & 250  &  300  &  300  &  300  &  250  &  250  &  500 \\
Telluric\footnote{Telluric standard stars used for the correction of the telluric absorption lines.} & $\rho$ Peg  &  c And  &  35 Vul  &  29 Vul  &  $\rho$ Peg  &  HR 8844  &  HR 196 \\
$N(\text{C}_2$)\footnote{The column densities of the C$_2$ molecules adopted from \citet{gre01}.}  (10$^{13}$ cm$^{-2}$) & --- &  10$^{+3.5}_{-1.5}$ &  $3.3$ &  5.2$\pm$1 &  --- &  $<3.3$ & 20$^{+4}_{-2}$ 
\enddata
\tablecomments{}
 \label{Cygtargets}
\end{deluxetable*}

\section{NIR DIBs toward Cyg OB2}

\begin{deluxetable*}{cccccccc}
\tabletypesize{\scriptsize}
 \tablecaption{EWs of DIBs}
\tablewidth{0pt}
\tablehead{
 \colhead{} & \colhead{No.\,3}  & \colhead{No.\,5}  & \colhead{No.\,8A}  & \colhead{No.\,9}  & \colhead{No.\,10}  & \colhead{No.\,11}  & \colhead{No.\,12} 
}
\startdata
$\lambda$9880    & 85.3 $\pm$ 6.1 & 59.5 $\pm$ 5.2 & 52.9 $\pm$ 4.2 & 53.8 $\pm$ 4.5 & 71.6 $\pm$ 6.9 & 77.0 $\pm$ 8.2 & 55.3 $\pm$ 4.7 \\
$\lambda$10360  & 77.0 $\pm$ 4.9 & 77.2 $\pm$ 3.8 & 94.8 $\pm$ 6.2 & 76.5 $\pm$ 5.4 & 97.9 $\pm$ 5.4 & 82.9 $\pm$ 4.8 & 74.0 $\pm$ 4.4 \\
$\lambda$10393  & 74.7 $\pm$ 6.5 & 46.3 $\pm$ 3.5 & 47.5 $\pm$ 4.8 & 48.6 $\pm$ 4.7 & 47.8 $\pm$ 4.2 & 48.6 $\pm$ 4.2 & 43.4 $\pm$ 4.8 \\
$\lambda$10438  & 89.4 $\pm$ 9.0 & 78.0 $\pm$ 4.5 & 125.8 $\pm$ 7.8 & 75.9 $\pm$ 5.8 & 134.3 $\pm$ 7.4 & 79.7 $\pm$ 7.4 & 92.7 $\pm$ 5.9 \\
$\lambda$10504  & 87.8 $\pm$ 2.4 & 73.1 $\pm$ 2.4 & 75.1 $\pm$ 2.7 & 83.7 $\pm$ 2.5 & 78.9 $\pm$ 3.8 & 71.7 $\pm$ 3.4 & 78.3 $\pm$ 2.4 \\
$\lambda$10780  & 403.2 $\pm$ 8.8 & $-$ & 240.3 $\pm$ 5.4 & 347.8 $\pm$ 5.9 & 486.3 $\pm$ 7.8 & 353.2 $\pm$ 8.5 & 313.6 $\pm$ 5.2 \\
$\lambda$10792  & 115.2 $\pm$ 5.9 & $-$ & 78.9 $\pm$ 5.5 & 85.9 $\pm$ 4.0 & 140.7 $\pm$ 8.1 & 95.6 $\pm$ 9.6 & 81.9 $\pm$ 6.1 \\
$\lambda$11797  & 347.8 $\pm$ 9.6 & 213.6 $\pm$ 8.3 & 220.6 $\pm$ 7.5 & 290.2 $\pm$ 6.8 & 401.3 $\pm$ 10.5 & 279.3 $\pm$ 10.3 & 270.0 $\pm$ 4.1 \\
$\lambda$12293  & 65.1 $\pm$ 5.7 & 51.1 $\pm$ 4.4 & 49.2 $\pm$ 4.9 & 54.1 $\pm$ 4.0 & 93.5 $\pm$ 5.9 & 60.8 $\pm$ 6.8 & 40.4 $\pm$ 3.5 \\
$\lambda$12337  & 239.1 $\pm$ 11.7 & 203.9 $\pm$ 6.2 & 129.1 $\pm$ 11.5 & 255.9 $\pm$ 7.4 & 208.2 $\pm$ 9.2 & 276.2 $\pm$ 19.2 & 270.5 $\pm$ 5.8 \\
$\lambda$12518  & 33.4 $\pm$ 5.4 & $-$ & 33.9 $\pm$ 4.8 & 29.2 $\pm$ 3.7 & $-$ & 29.2 $\pm$ 5.9 & 41.5 $\pm$ 4.2 \\
$\lambda$12536  & 78.1 $\pm$ 5.9 & 57.9 $\pm$ 6.4 & 73.5 $\pm$ 6.1 & 89.7 $\pm$ 5.5 & 92.7 $\pm$ 7.2 & 74.4 $\pm$ 7.8 & 54.5 $\pm$ 4.9 \\
$\lambda$12623  & 245.5 $\pm$ 10.0 & 135.0 $\pm$ 10.6 & 110.7 $\pm$ 7.1 & 185.7 $\pm$ 7.6 & 183.4 $\pm$ 10.3 & 140.9 $\pm$ 11.2 & 159.0 $\pm$ 4.1 \\
$\lambda$12799  & 158.0 $\pm$ 6.2 & 118.6 $\pm$ 6.8 & 62.3 $\pm$ 7.5 & 89.0 $\pm$ 4.5 & 78.5 $\pm$ 4.5 & 122.1 $\pm$ 9.2 & 94.8 $\pm$ 2.8 \\
$\lambda$12861  & 97.1 $\pm$ 8.8 & 56.6 $\pm$ 4.8 & 79.4 $\pm$ 10.6 & 97.0 $\pm$ 5.7 & 130.5 $\pm$ 9.9 & 109.6 $\pm$ 8.3 & 77.6 $\pm$ 3.8 \\
$\lambda$13027  & 155.3 $\pm$ 7.6 & 84.9 $\pm$ 7.9 & 145.5 $\pm$ 6.1 & 131.3 $\pm$ 5.7 & 240.6 $\pm$ 8.6 & 115.9 $\pm$ 8.3 & 106.0 $\pm$ 4.8 \\
$\lambda$13175  & 1374.8 $\pm$ 16.8 & 949.8 $\pm$ 14.6 & 905.9 $\pm$ 12.9 & 1197.3 $\pm$ 14.0 & 1475.7 $\pm$ 15.8 & 1270.0 $\pm$ 18.5 & 1091.2 $\pm$ 9.0
\enddata
\tablecomments{The EWs are given in units of m\r{A}. The bars mean that the EW or upper limit of the DIB could not be evaluated due to the overlapped stellar absorption lines and/or the residual features of the telluric correction.}
 \label{CygDIBew1}
\end{deluxetable*}

\begin{figure*}
 \centering
 \includegraphics[width=15cm,clip]{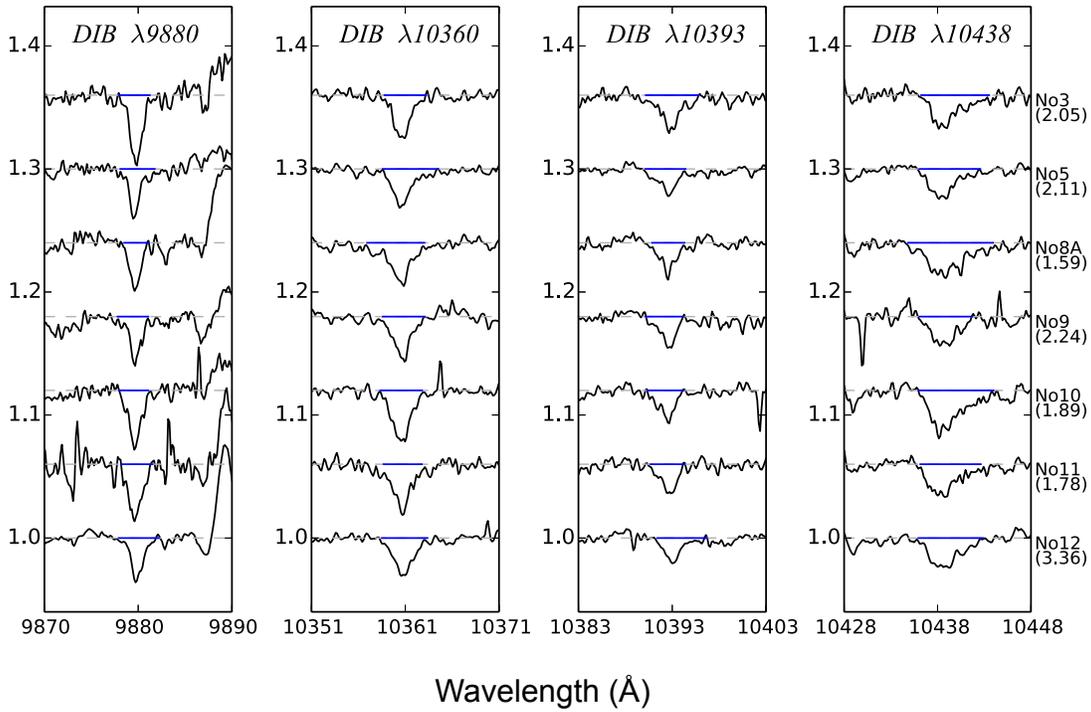}
 \caption{Spectra of four DIBs ($\lambda \lambda 9880, 10360, 10393, \text{and }10438$) from all targets. The names and $E(B-V)$ values of the targets are displayed along the right side of the $\lambda$10438 plot. The spectra are normalized and plotted with arbitrary offsets. Thin dashed line is the continuum level of each star. When the EWs or upper limits of the DIBs cannot be evaluated because their stellar and/or telluric absorption lines overlap, the spectra are plotted as gray lines. Thick blue lines are the integrated ranges used in the EW calculations.}
 \label{Cyg_DIBspec1}
\end{figure*}

\begin{figure*}
 \centering
 \includegraphics[width=15cm,clip]{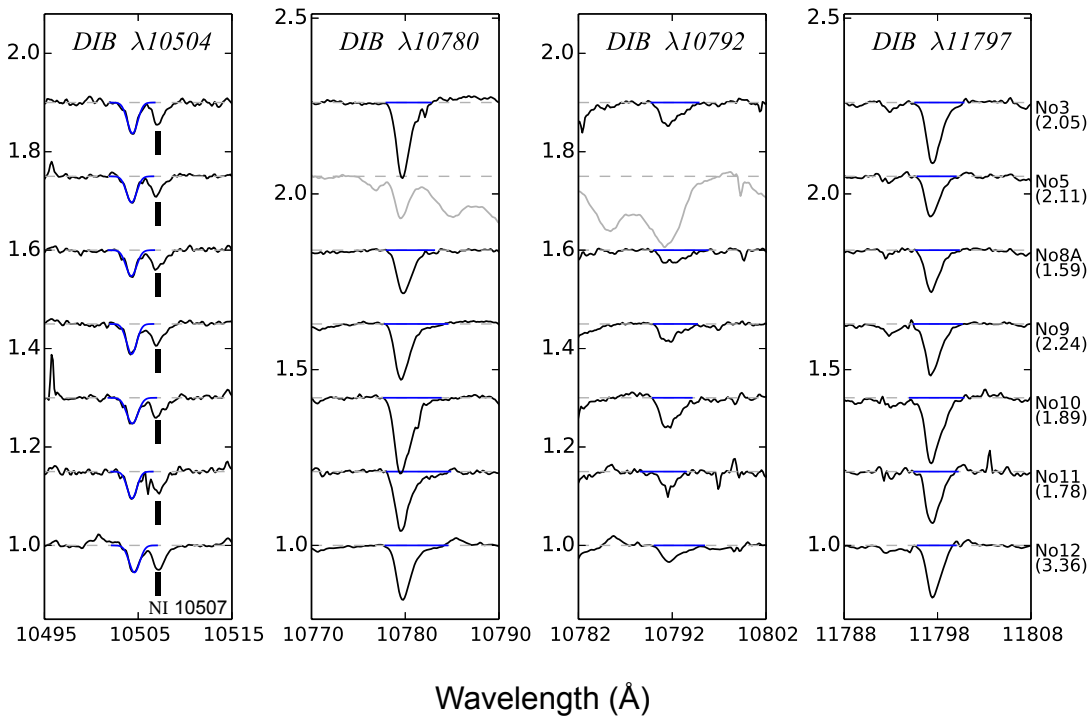}
  \caption{Spectra of four DIBs ($\lambda \lambda$10504, 10780, 10792, and 11797) from all targets.  The notations are specified in Figure \ref{Cyg_DIBspec1} except for DIB $\lambda 10504$, whose EWs are estimated by fitting \textcolor{black}{Gaussian profiles to the spectra} (blue lines) to avoid blending of the stellar absorption lines of \ion{N}{1} 10507. }
 \label{Cyg_DIBspec2}
\end{figure*}

\begin{figure*}
 \centering
 \includegraphics[width=15cm,clip]{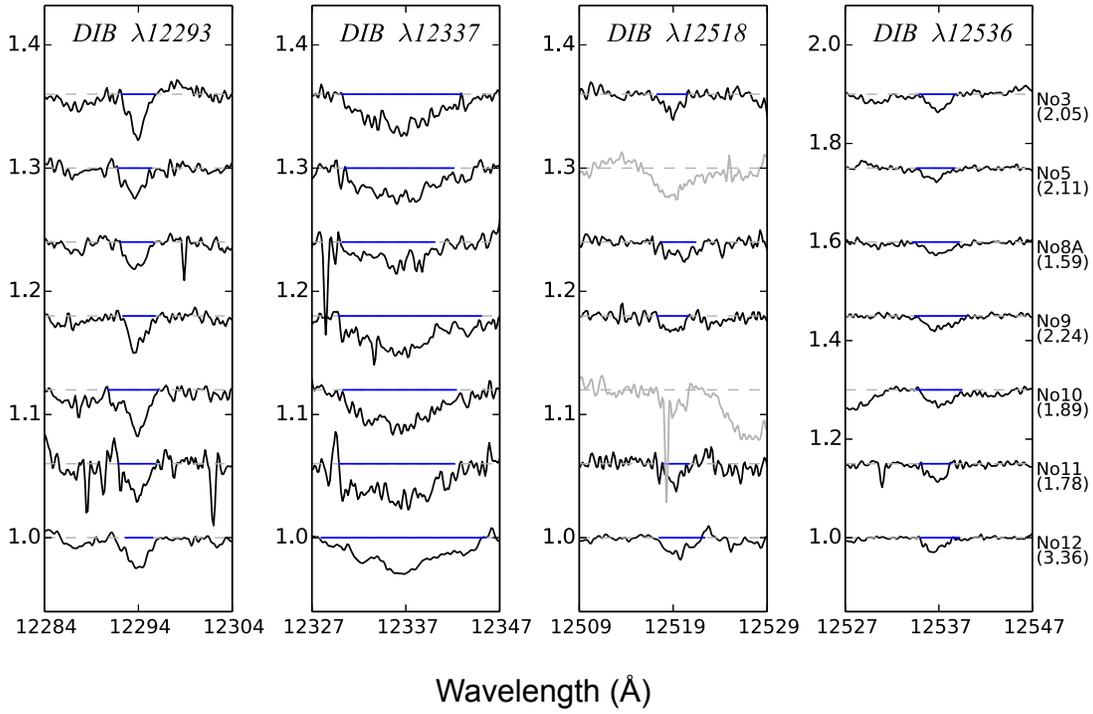}
 \caption{Spectra of four DIBs ($\lambda \lambda$12293, 12337, 12518, and 12536) from all targets. The notations are specified in Figure \ref{Cyg_DIBspec1}. \textcolor{black}{[4] Although $\lambda$12518 is detected in the spectrum of No.\,5, the EW of $\lambda$12518 cannot be estimated because the neighboring continuum is affected by the strong \ion{He}{1} 12527 line.}}
 \label{Cyg_DIBspec3}
\end{figure*}

\begin{figure*}
 \centering
  \includegraphics[width=15cm,clip]{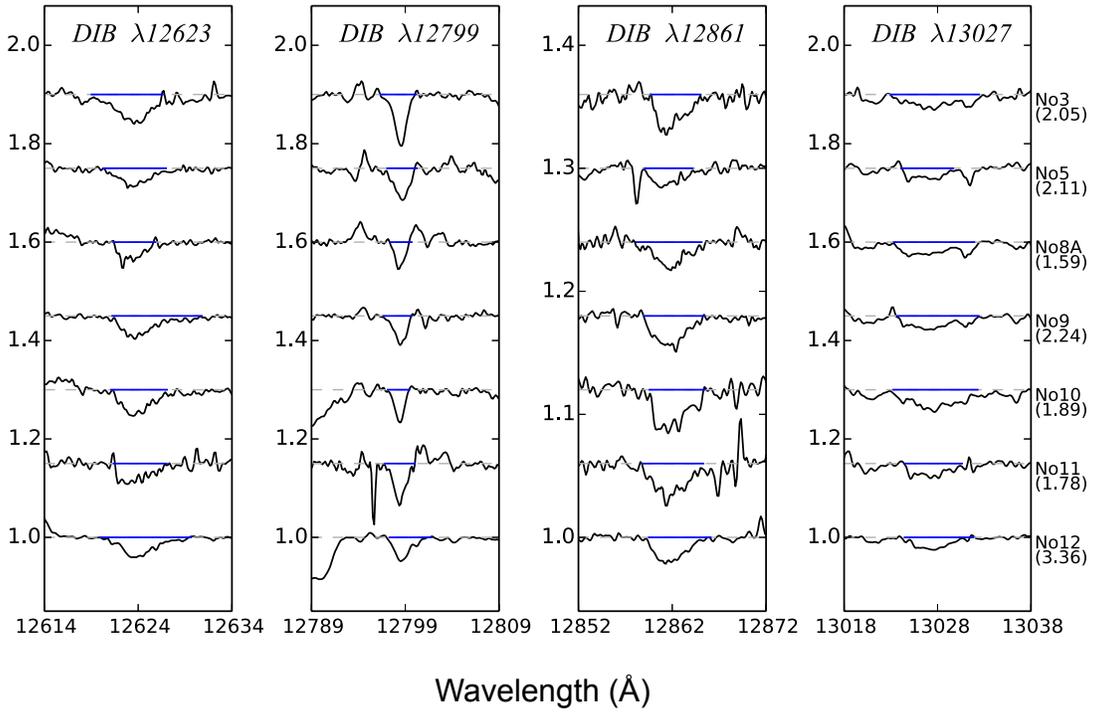}
 \caption{Spectra of four DIBs ($\lambda \lambda$12623, 12799, 12861, and 13027) from all targets. The notations are specified in Figure \ref{Cyg_DIBspec1}. }
 \label{Cyg_DIBspec4}
\end{figure*}

\begin{figure}
 \centering
  \includegraphics[width=4cm,clip]{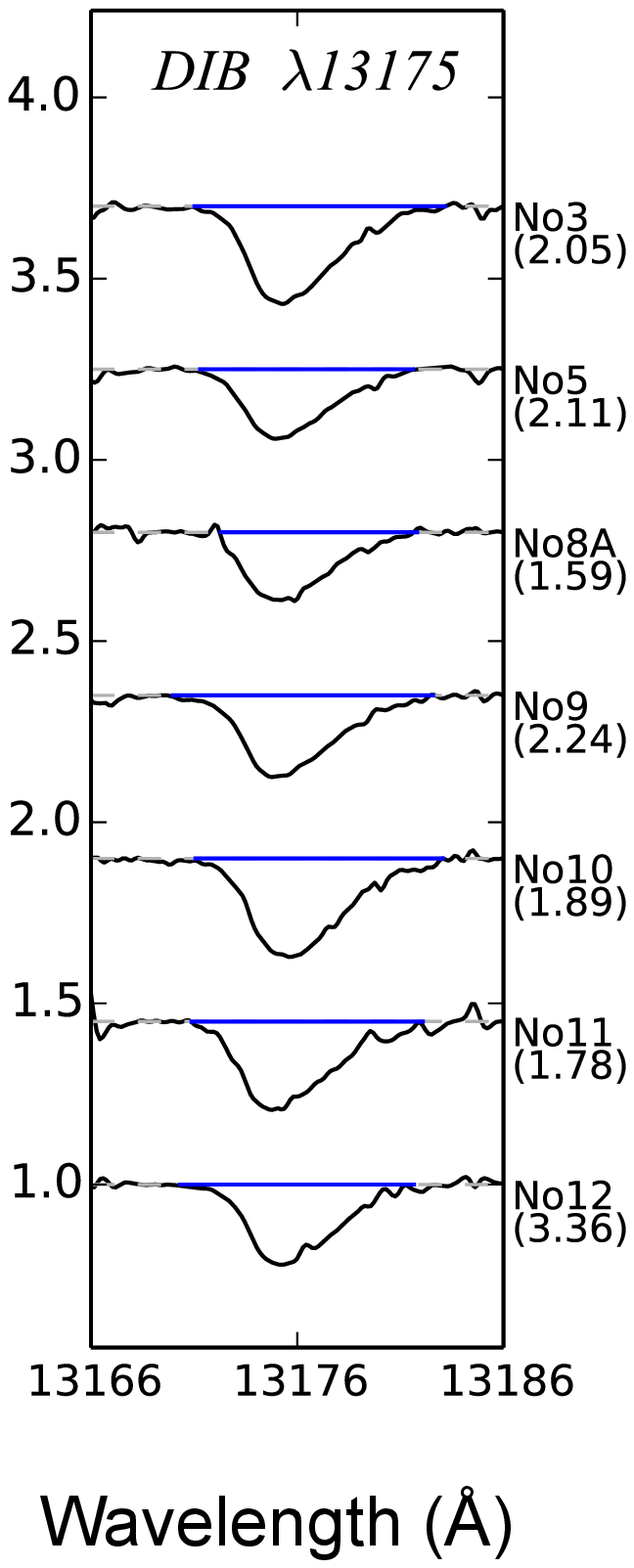}
 \caption{Spectra of DIB $\lambda 13175$ from all targets. The notations are specified in Figure \ref{Cyg_DIBspec1}. }
 \label{Cyg_DIBspec5}
\end{figure}

In Paper I, we analyzed 5 previously known DIBs and 15 newly identified ones. Therefore, we searched for these 20 DIBs in the obtained spectra of the seven stars in Cyg OB2. Owing to the large EWs and high S/N of the spectra, all DIBs were clearly detected toward all the observed stars. However, we removed three DIBs ($\lambda \lambda9577, 9633$, and 10697) because their stellar absorption lines overlap, leaving 17 DIBs in the following analysis. Figures \ref{Cyg_DIBspec1} --\ref{Cyg_DIBspec5} show the DIB spectra of all observed stars. As in Paper I, we calculated the EWs by a simple integration because the DIB profiles are not well established. The exception was DIB $\lambda 10504$, whose EWs cannot be evaluated by simple integration because of blending with stellar lines. In this case, we estimated the EWs by Gaussian-fitting to the DIB profile. The wavelength range of the integration was arbitrarily set to include the whole DIB profile. \textcolor{black}{[3] The uncertainties of the EWs ($\sigma _{\text{err}}$) were estimated from the S/N of the neighboring continuum per pixel by the following equation: }
\begin{equation}
\sigma _{\text{err}}  = \Delta \lambda \sqrt{w} (\text{S/N})^{-1}, \\
\end{equation}
\textcolor{black}{where $\Delta \lambda$ is the wavelength interval per resolution elements ($ \sim 0.4$ \r{A}), $w$ is the width of the DIB in resolution elements.} Additional systematic uncertainties introduced by continuum fitting, telluric absorption lines, and stellar lines were ignored. Table \ref{CygDIBew1} summarizes the EWs of the NIR DIBs and their uncertainties. The EWs of $\lambda \lambda 10780$, 10792, and 12518 toward No.\,5 could not be estimated because the stellar absorption lines overlap on the DIBs. The EWs of $\lambda$12518 toward No.\,10 were also inestimable because of a spurious feature.

\section{Correlations of NIR DIBs toward Cyg OB2}

In Paper I, we investigated the correlations among the NIR DIBs, some strong optical DIBs, and the reddening of stars. Consequently, we suggested two "families" of well-correlated DIBs. We suggested that the NIR DIBs in \textcolor{black}{a} family have similar physical properties or are chemically related to each other. The gas clouds toward most of the targets in Paper I (on average) are expected to reside in a uniform interstellar environment. Compared to the targets in Paper I, the lines of sight of Cyg OB2 are considered to reside in unique environments in view of their density structure and UV flux. Here we investigate the environmental dependence of these DIBs from their correlations in the Cyg OB2 environment. 

\subsection{An NIR DIB family}

\begin{figure*}
 \centering
 \includegraphics[width=16cm,clip]{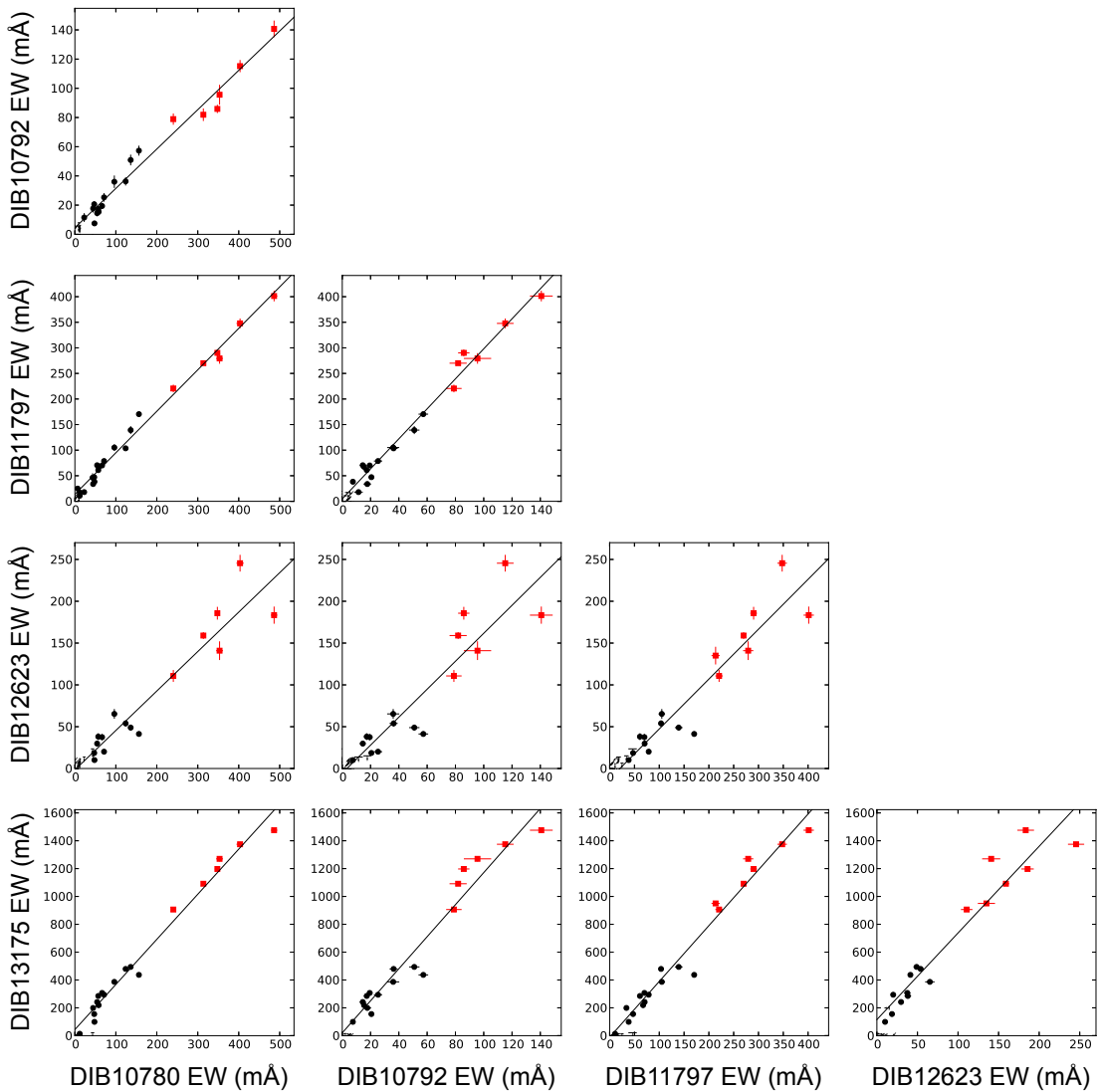}
 \caption{Correlations among the \textcolor{black}{five} NIR DIBs, $\lambda \lambda$10780, 10792, 11797, 12623, and 13175. Black and red points indicate the data from Paper I and from the stars in Cyg OB2, respectively. The data in each panel were fitted \textcolor{black}{with} linear functions (straight lines in the plots).}
 \label{F1_Cyg}
\end{figure*}

In Paper I, we suggested two families of NIR DIBs: 1) $\lambda \lambda$10780, 10792, and 11797 as well as 2) $\lambda \lambda$10504, 12623, and 13175. Figure \ref{F1_Cyg} shows the correlations among the DIBs in each family, excluding $\lambda 10504$, which will be discussed in the next subsection. The strong correlations in Paper I appear also in Cyg OB2. Table \ref{correlations} lists the correlation coefficients calculated from the EWs in Paper I and in Cyg OB2. The correlation coefficients among the DIBs are comparable in Paper I and the present Cyg OB2 study. In addition, no distinct difference between the two families is evident. The five DIBs seem to be highly correlated, suggesting that they compose one ``family.'' Note that the correlations between $\lambda 12623$ and the other four DIBs are more dispersed than the other paired correlations (Fig. \ref{F1_Cyg}); thus, their correlations coefficients are lower than the other pairs (Table \ref{correlations}). The classifications of the two families in Paper I were based on the low correlation coefficients between the groups $\lambda \lambda$10504 and 12623 as well as $\lambda \lambda$10780, 10792, and 11797. Because the DIBs were weaker than those toward Cyg OB2, the correlation coefficients may have been affected by systematic uncertainties of the DIB EWs introduced by contaminated stellar absorption lines and telluric absorption lines. The large EWs of the DIBs toward Cyg OB2 clarify the correlations among the DIBs. The very tight correlations in the Cyg OB2 environment suggest some similarity in the physical and/or chemical properties of the DIB carriers of the five \textcolor{black}{bands}. \textcolor{black}{[5] However, we should keep in mind that the correlations for the lines of sight with large extinction may naturally become better because the difference of the ratios of the DIB EWs among the multiple clouds are averaged to be smeared out \citep{bai15a}. It is impossible to resolve the multiple velocity components from the observed DIB profiles due to the intrinsically large width and the lack of the information on the intrinsic
     profile. For more convincing study of the DIB correlations, it is
     necessary to observe the single-cloud lines of sight \citep{cam97} and/or a number of lines of sight \citep{fri11}.}

\begin{deluxetable}{cccccc}
\tabletypesize{\scriptsize}
 \tablecaption{Correlation coefficients of NIR DIBs}
\tablewidth{0pt}
\tablehead{
 \colhead{} & \colhead{$\lambda10780$}  & \colhead{$\lambda10792$}  & \colhead{$\lambda11797$}  & \colhead{$\lambda12623$}  & \colhead{$\lambda13175$}  
}
\startdata
$\lambda$10780& $-$ & $-$& $-$ & $-$ & $-$\\ 
$\lambda$10792& 0.99 (19) & $-$ & $-$ & $-$ & $-$\\ 
$\lambda$11797& 0.99 (23) & 0.99 (19) & $-$ & $-$ & $-$\\ 
$\lambda$12623& 0.95 (16) & 0.91 (16) & 0.94 (17) & $-$ & $-$\\ 
$\lambda$13175& 0.99 (19) & 0.97 (18) & 0.98 (20) & 0.96 (17) & $-$ 
\enddata
\tablecomments{The number in parentheses denotes the number of stars used for the calculation of the correlation coefficients. The data of stars observed in both Paper I and this paper are included in this calculation.}
 \label{correlations}
\end{deluxetable}

\begin{figure*}
 \centering
 \includegraphics[width=13cm,clip]{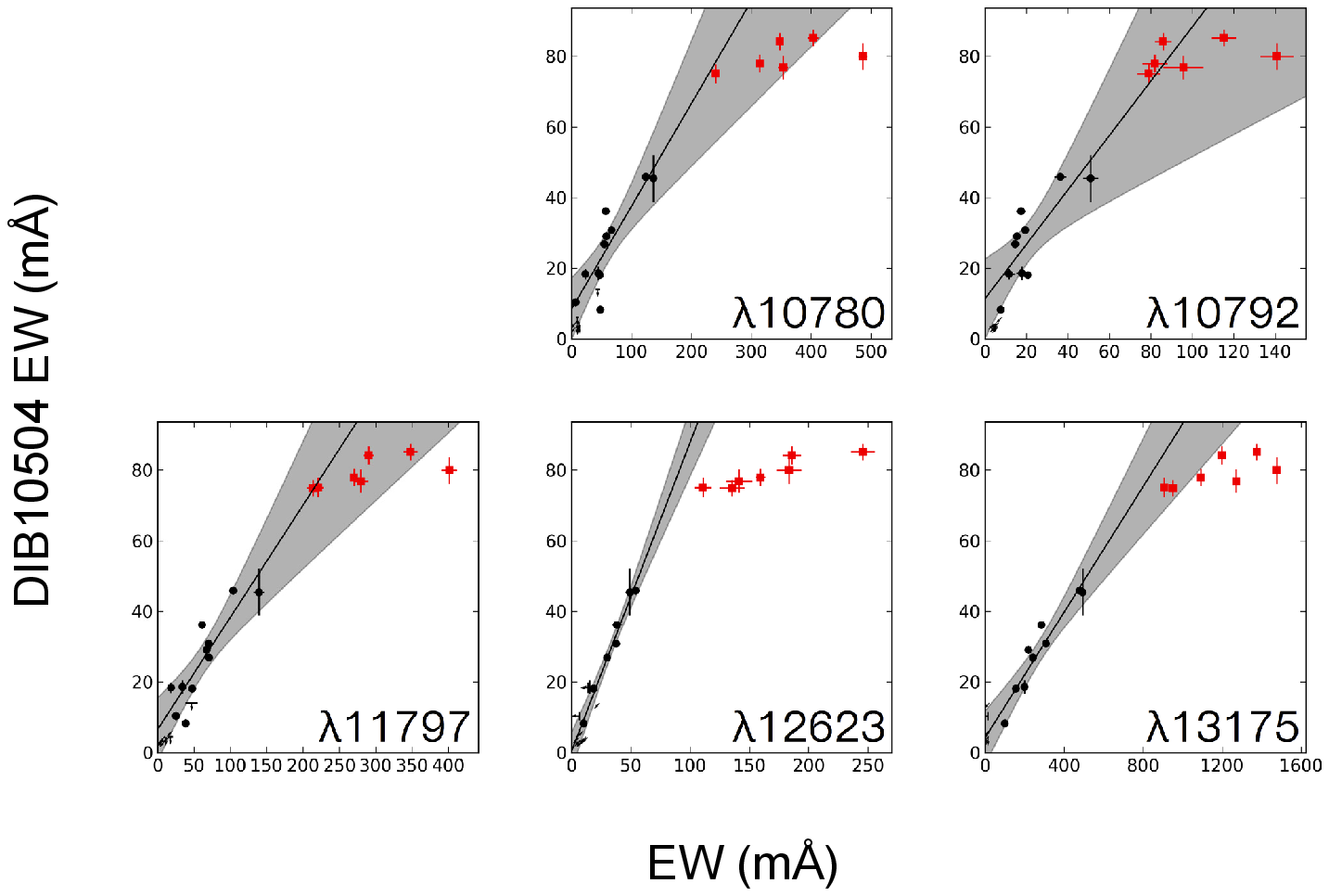}
 \caption{Correlations of $\lambda 10504$ versus the family DIBs ($\lambda \lambda$10780, 10792, 11797, 12623, and 13175). The lines in each panel \textcolor{black}{are} linearly fitted to the black points in each plot. The shaded areas represent the \textcolor{black}{95\% confidence bands.}}
 \label{10504_Cyg}
\end{figure*}

\subsection{$\lambda$10504: not a member of the family}

In Paper I, we suggested that $\lambda 10504$ is also well correlated with $\lambda \lambda 12623$ and 13175, although the correlation coefficients were calculated from relatively few data points. Figure \ref{10504_Cyg} shows the correlations between $\lambda$10504 and the family NIR DIBs. We also plot the least-squares regression lines calculated from the data in Paper I and \textcolor{black}{the 95\% confidence bands}. The present correlations differ from those in Paper I; in particular, \textcolor{black}{[6] the ratios of the EW of $\lambda 10504$ to the EWs of the other DIBs in Cyg OB2 are smaller than those in the diffuse interstellar environment.} Moreover, the EWs of $\lambda$10504 are almost constant across Cyg OB2, whereas those of the DIB family vary by a factor of two. \textcolor{black}{[8] Note that the EWs of $\lambda 10504$ in Cyg OB2 appears to be better correlated with those of $\lambda 12623$ (correlation coefficient is $\sim 0.9$) than with those of the other four family DIBs (correlation coefficients are $\leq 0.7$), as is the same in diffuse interstellar environment.}

The smaller EWs of $\lambda$10504 relative to the DIB family in Cyg OB2 could be explained by saturation of \textcolor{black}{the} $\lambda$10504's absorption lines, which would cause the slower increase in the EWs of $\lambda$10504 than in the EWs of the other DIBs. Assuming full saturation of the absorption lines, the average EW of $\lambda$10504 in Cyg OB2 is 78.4 $\pm$ 5.4 m\r{A}, corresponding to a velocity width $\Delta v \sim 2$ km s$^{-1}$. Because this velocity width should include the intrinsic DIB width, the velocity dispersion of the multiple lines-of-sight components and the Doppler widths of the various velocity components, the calculated value of $\sim$ 2 km s$^{-1}$ appears to be unrealistically small. Therefore, we conclude that $\lambda$10504 is not saturated, and that its EWs are decreased relative to the family by the column density of $\lambda 10504$ carrier (alternatively, the EWs of the family are increased relative to $\lambda$10504) along the line of sight of Cyg OB2.

\section{Discussion}

\subsection{NIR DIBs correlations with C$_2$}

Because C$_2$ is the simplest form of the carbon-based molecules, its relationship with DIBs is important for constraining the carriers. 
\citet{tho03} showed that some weak DIBs (such as $\lambda \lambda$4936.96 and 5418.91) are well correlated with the column densities of C$_2$ molecules, suggesting that their carriers are chemically related to C$_2$. \citet{ada05} showed that the carriers of the so-called ``C$_2$ DIBs'' are abundant in the cores of diffuse clouds, where the molecular fraction of hydrogen is high and some strong DIBs (e.g., $\lambda \lambda$5780, 5797 and 6614) are weak. They suggested that the C$_2$ DIB carriers are distinct from the carriers of the strong DIBs, and occupy the denser cloud regions than the strong DIBs. Therefore, by correlating the DIBs with C$_2$ molecules, we can elucidate the density dependence of DIBs. Especially at NIR wavelengths, we can examine the DIBs in molecular clouds \citep{ada94}.

\begin{figure*}
 \centering
 \includegraphics[width=15cm,clip]{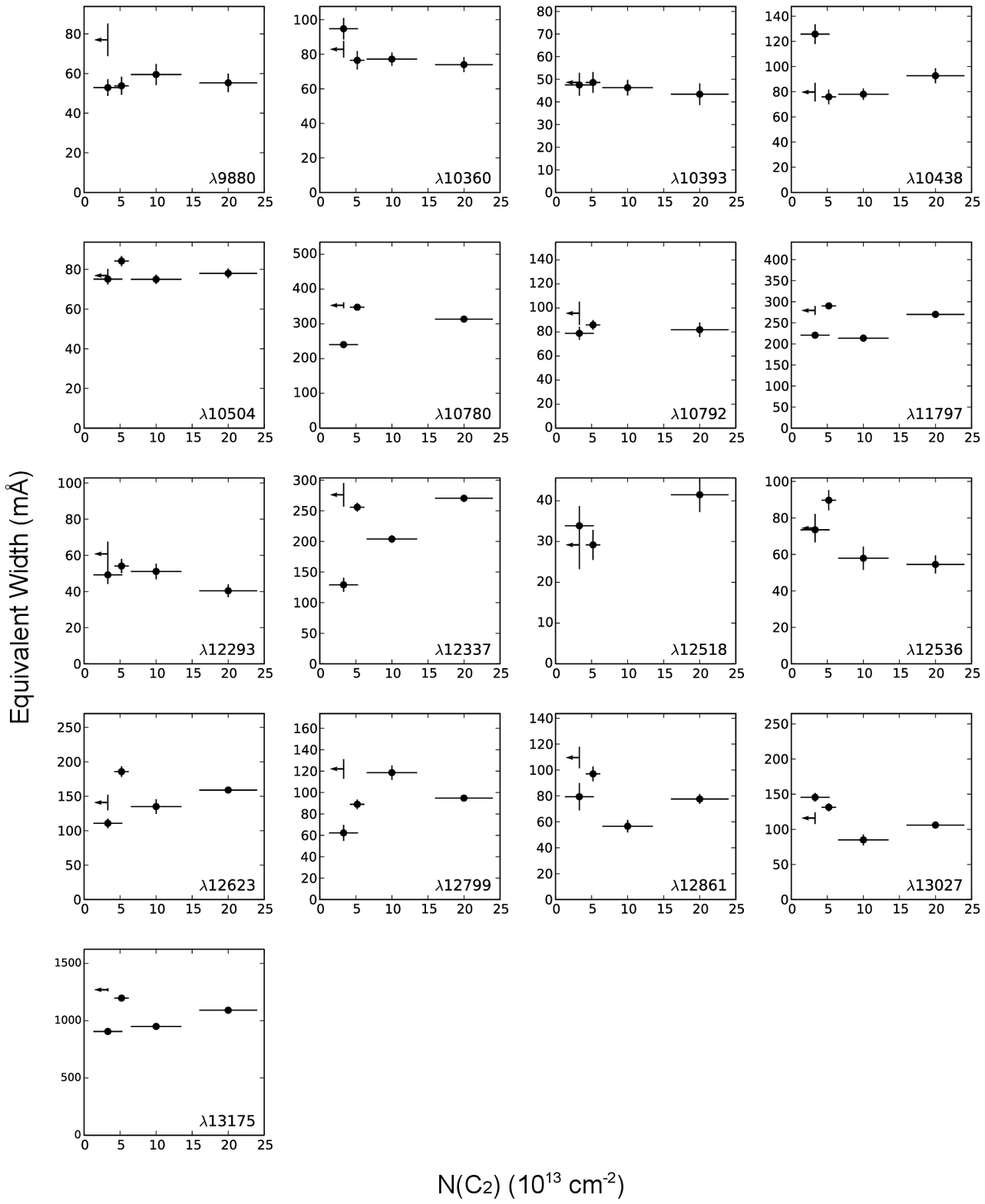}
 \caption{EWs of the 17 NIR DIBs as functions of the column density of C$_2$ molecules. The vertical and horizontal axes represent the EW of the NIR DIBs and the C$_2$ column densities, respectively, adopted from \citet{gre01} (see also Table \ref{Cygtargets}).}
 \label{Cyg_cc_C2}
\end{figure*}

The gas clouds toward Cyg OB2 comprise both diffuse and dense components. By correlating the DIBs against the C$_2$ molecules toward Cyg OB2, we can study whether the carriers of all 17 NIR DIBs are distributed in the dense component. From the C$_2$ (2,0) Phillips band around $\lambda =7720$\r{A} \citep{gre01}, 
the column densities of C$_2$ toward No.\,5, 7, 8A, 9, 11, and 12 were estimated as 10$^{+3.5}_{-1.5}$, $<3.3$, 3.3, $5.2\pm 1$, $<3.3$ and $20^{+4}_{-2}$ $\times 10^{13} \text{ cm}^{-2}$, respectively. Figure \ref{Cyg_cc_C2} shows the correlations between the NIR DIBs and C$_2$ column densities toward Cyg OB2. The column densities of C$_2$ toward Cyg OB2 widely vary, and all NIR DIBs are fairly strong along all observed lines of sight. Although only five lines of sight are available, none of the NIR DIBs are obviously correlated with C$_2$. This suggests that the NIR DIBs are mainly contributed by the diffuse rather than the dense cloud component. We propose that all 17 of the NIR DIBs are distinct from ``C$_2$ DIBs''.

\subsection{DIB $\lambda10504$ as an environment-sensitive line}

\begin{figure*}
 \centering
  \includegraphics[width=14cm,clip]{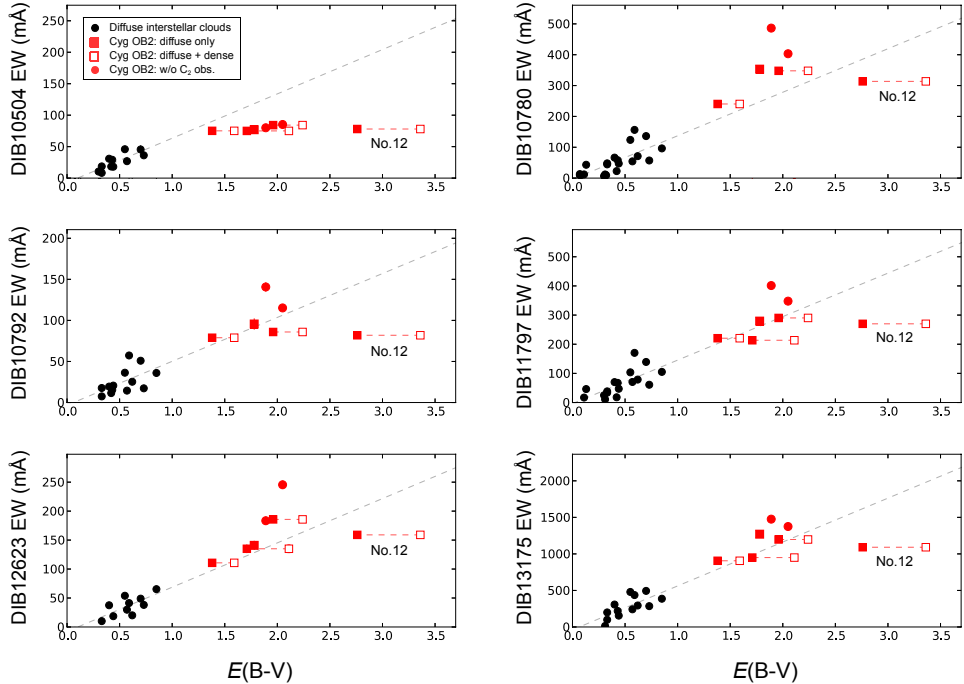}
 \caption{\textcolor{black}{EWs of $\lambda \lambda$10504, 10780, 10792, 11797, 12623, and 13175 as a function of $E(B-V)$. Red open squares show the EWs of the OB stars in Cyg OB2 with and without C$_2$ detection \citep{gre01}. Red circles show the EWs of DIBs toward No.\,3 and No.\,10, the directions of no reported C$_2$ absorption bands. The EWs of the lines of sight with C$_2$ detection are plotted using $E(B-V)$ contributed only from the diffuse clouds with filled marks. The marks for the same stars are connected with the red dashed lines. The gray dashed lines are the best linear fitting for the black points derived in Paper I.}}
  \label{ebv_corr}
\end{figure*}

In \textsection{4}, $\lambda \lambda$10780, 10792, 11797, 12623, and 13175 were found to constitute a family, whereas $\lambda$10504, which was originally suggested to be well correlated with this family, is not well correlated with the same DIBs in Cyg OB2. $\lambda$10504 would differ from the family if the $\lambda$10504 carrier decreased and/or the carriers of the family NIR DIBs increased. To elucidate the cause of this difference, we examined the $E(B-V)$ dependence of DIB $\lambda$10504 \textcolor{black}{and the family NIR DIBs}. 
Figure \ref{ebv_corr} \textcolor{black}{shows the plots of EW -- $E(B-V)$ for $\lambda$10504 and the family NIR DIBs}. We also plot the least-square regression lines derived in Paper I. 

\textcolor{black}{[7] In \textsection{5.1}, we suggested that the DIB EW is contributed only by diffuse components. On the other hand, the $E(B-V)$ is contributed by both diffuse and dense components. }
\textcolor{black}{To discuss the DIB EWs as a function of $E(B-V)$ contributed only from diffuse clouds, we estimate the $E(B-V)$ of dense component from the C$_2$ column densities. First, the H$_2$ column densities are estimated using the following $\log N(\text{H}_2)$ -- $\log N(\text{C}_2)$ correlation derived for the translucent lines of sight by \citet{son07}\footnote{\textcolor{black}{Because they showed only the slope, we calculated the intercept from their plot of $\log N(\text{H}_2)$ -- $\log N(\text{C}_2)$ \citep[see Figure 17 of ][]{son07}.}}:}
\begin{equation}
\log N(\text{C}_2) = 1.79 \log N(\text{H}_2) - 23.7 \label{c2-h2}
\end{equation}
\textcolor{black}{The C$_2$ and H$_2$ column densities are well correlated as far as C$^+$ is the dominant form of gas-phase carbon atoms because C$^+$ and H$_2$ are necessary for the formation of C$_2$ molecules by the interstellar chemistry \citep{van89,van91,son07}. Subsequently, the estimated H$_2$ column densities are converted to $E(B-V)$ using the following relation \textcolor{black}{for diffuse interstellar clouds} \citep{boh78}:}
\begin{equation}
\left\{ N(\text{\ion{H}{1}}) +  2 N(\text{H}_2) \right\} / E(B-V) = 5.8 \times 10^{21} \text{cm}^{-2} \text{mag}^{-1} ,
\end{equation}
\textcolor{black}{where we assumed $N(\text{\ion{H}{1}}) = 0$. The estimated $E(B-V)$ of dense component are summarized in Table \ref{Ebvcorrection}. The $E(B-V)$ of diffuse component toward each star are calculated by subtracting $E(B-V)$ of dense component from $E(B-V)$. Here, we assume that all the dense components can be traced only with C$_2$ and the column densities toward the stars of Cyg OB2 satisfy the relation (\ref{c2-h2}). In Fig. \ref{ebv_corr}, we also plot the points using the resultant $E(B-V)$ of diffuse component (filled squares). As a result, the points of stars with C$_2$ detection, except for No.\,12, appear to be consistent with the point of the star without C$_2$ detection, No.\,11. Only No.\,12, which is the most reddened star in Cyg OB2, is still located much lower than the regression lines for all DIBs in Fig. \ref{ebv_corr}. We will discuss this exceptional line of sight at the end of this subsection.}

\textcolor{black}{The EWs of $\lambda 10504$ toward \textit{all} observed stars in Cyg OB2 are smaller than that expected from the correlation; in fact, they are \textit{almost constant} across Cyg OB2, consistent with the DIB - DIB correlations in Figure \ref{10504_Cyg}. 
On the other hand, the family DIBs, except for $\lambda$12623, appear to be almost consistent with the regression lines except for No.\,12.} \textcolor{black}{ Therefore, we conclude that the difference seen in Figure \ref{10504_Cyg} is mainly caused by the decrease of the $\lambda 10504$ carrier rather than the increase of the carriers of the family NIR DIBs. }

\textcolor{black}{Only for DIB $\lambda$12623, the EWs are found to be stronger than expected from the regression line (except for No.\,12) and rather appear to be systematically increasing as a function of $E(B-V)$ of diffuse component. The EWs of $\lambda$12623 in Cyg OB2 may have different dependence on $E(B-V)$ from $\lambda\lambda$10780, 10792, 11797 and 13175 because the correlation coefficients between $\lambda$12623 and the four family DIBs are slightly smaller than those among the four family DIBs (Table \ref{correlations}). However, it is necessary to increase the sample size to examine the possibility.} \textcolor{black}{Note that $\lambda$10780 EWs of Cyg OB2 for all objects but No.\,12 also appear to be larger than expected from the regression line. However, the difference of the DIB EWs from the regression line is not statistically significant in view of the small number of points used to derive the regression lines. Considering the tight correlations between $\lambda$10780 and $\lambda\lambda$10792, 11797, and 13175 (Fig. \ref{F1_Cyg}), it is unlikely that only $\lambda$10780 has a different dependence on $E(B-V)$ in the Cyg OB2 environment.}

\textcolor{black}{[9] In the studies of environmental tendency of optical DIBs, the EWs of some DIBs (e.g., $\lambda$6284) were also nearly constant as a function of the extinction (or EW$/E(B-V)$ $\propto E(B-V)^{-1}$) in well-known dense regions, such as Orion molecular clouds \citep{jen94a}, $\rho$ Ophiuchi cloud complex, and Taurus dark clouds \citep{ada91}. The dependence is suggested to be caused by the lower abundance of DIB carriers in the dense clouds, which are likely to have high $E(B-V)$ values. Such property is also seen in all NIR DIBs by the lack of any correlation with the C$_2$ column densities (see \textsection{5.1}). Although the EW dependence of $\lambda$10504 of Cyg OB2 appears to be the same as the tendency in the dense regions at a first look, the cause is totally different because the carrier of $\lambda$10504 is not present} \textcolor{black}{even in the diffuse component of Cyg OB2.}

\citet{wri15} calculated the extinction distribution of OB stars in Cyg OB2 as $A_V =$ 4--7 mags (median $A_V = 5.4$ mag). 
According to the regression line in Figure \ref{ebv_corr}, the average $\lambda 10504$ EW (78.4 $\pm$ 5.4 m\r{A}) corresponds to $E(B-V) \sim 1.2$ mag ($\sim$ 3.6 mag in $A_V$), similar to the lower limit of the extinction distribution. This suggests that the $\lambda10504$ carrier is not distributed in the diffuse component \textcolor{black}{of} Cyg OB2 but exists only in foreground diffuse clouds, such as the Cygnus Rift \citep{gua12}. If the foreground interstellar clouds along the lines of sight uniformly cover the association, the EWs would be almost constant (as observed). The environmental dependence of DIB $\lambda 10504$ may be attributed to the strong flux of the ionizing photons from early-type stars in Cyg OB2. \textcolor{black}{[10] Such decrement of DIB absorptions in the harsh environment is also reported by \citet{van13} for DIBs $\lambda \lambda$4428 and 6614 near the OB associations within the Tarantula nebula.} We suggest that these ionizing photons in diffuse clouds completely destroy the $\lambda$10504 carrier. 
The different behaviors from the family NIR DIBs are thus attributable to different stabilities of the carrier molecules. 

\textcolor{black}{In Fig. \ref{ebv_corr}, the EWs of No.\,12 is much lower than expected from the regression lines even for the $E(B-V)$ of diffuse component. No.\,12 is a luminous blue variable (LBV) candidate and known to be one of the most intrinsically luminous star in the Galaxy \citep{hum78,cla12}. The extinction of this star ($A_V = 10.2$ mag) is extremely high compared to the other members of the cluster \citep[$4 < A_V < 7$ mag; ][]{wri15}. The cause of the extinction is not fully understood. Considering the extinction distribution of the early-type stars in Cyg OB2, the excess of the extinction must be $A_V > 3$ mag. In view of the extinction gap between the EWs of family NIR DIBs and the regression lines in Fig. \ref{ebv_corr}, the color excess from dense component is roughly estimated as $E(B-V) \sim 1.5$ mag ($\sim$ 5 mag in $A_V$). \citet{whi15} suggested that the excess of the extinction is caused by two or more translucent clouds because the ice absorption, which is expected for a single cloud of $A_V \geq 5$ mag, has not been detected in the spectrum of No.\,12.\footnote{\textcolor{black}{According to the recent study by \citet{mar16}, there is also a possibility that No.\,12 is truly an LBV and a part of the extinction ($A_V \sim 1$ mag) is caused by its circumstellar shell. Even if the possible contribution of the circumstellar shell is considered together with the extinction from the translucent clouds, the extinction gap between the EWs of family NIR DIBs and the regression lines in Fig. \ref{ebv_corr} cannot be explained well.}} In fact, the absorption lines of CO and C$_2$ in the spectrum of No.\,12 are detected in association with two velocity components at $v_{\text{LSR}} \sim 7, 12$ km s$^{-1}$ \citep{sca02,mcc02,cas05}. Because a large amount of carbon is transformed into CO in the dense regions of the translucent clouds, the C$_2$ absorption bands trace the H$_2$ molecules only in the surface. In fact, the density ranges traced by C$_2$ and $^{13}$CO are not the same \citep{mcc02}. Therefore, $E(B-V)$ contributed from dense component for No.\,12 may not be able to be estimated only by the C$_2$ column density. \citet{whi15} estimated the extinction of molecular clouds from $N(^{13}\text{CO})$ toward this star as $A_V \sim 1$ mag, which is still less than the excess, suggesting that the high UV intensity results in molecular abundances lower than those are typical for a given density regime. Due to the relatively complex structure of the line-of-sight clouds compared to the other stars, the extinction of No.\,12 contributed from the translucent clouds cannot be estimated straightforwardly.}

\begin{deluxetable*}{ccccc}
\tabletypesize{\scriptsize}
 \tablecaption{$E(B-V)$ correction}
\tablewidth{0pt}
\tablehead{
\colhead{Stars} & \colhead{$E(B-V)$} & \colhead{$N(\text{C}_2)$\footnote{The column densities of the C$_2$ molecules adopted from \citet{gre01}.}} & \colhead{$E(B-V)_{\text{dense}}$} & \colhead{$E(B-V)_{\text{diffuse}}$} \\
\colhead{} & \colhead{} & \colhead{$(10^{13} \text{cm}^{-2})$} & \colhead{} & \colhead{}
}
\startdata
No.\,3 & 2.05  &  ---  &  --- &  --- \\
No.\,5 & 2.11  & 10  &  0.40 &  1.71 \\
No.\,8A & 1.59  & 3.3 & 0.21 &1.38 \\
No.\,9 & 2.24  & 5.2 & 0.28 &  1.96 \\
No.\,10 & 1.89  &  ---  &  --- &  ---  \\
No.\,11 & 1.78  &  $<3.3$  &  0.0 &  1.78  \\
No.\,12 & 3.36  &  20 &  0.60  &  2.76
\enddata
\tablecomments{}
 \label{Ebvcorrection}
\end{deluxetable*}

\subsection{Implications for the NIR DIB carriers}

In Paper I, we suggested that the family of NIR DIBs arise from cations, because the DIBs are better correlated with $\lambda$5780.5 than with $\lambda$5797.1 and become stronger along the $\sigma$-type lines of sight (tracing the surface layer of diffuse clouds with strong UV fields) than along the $\zeta$-type lines of sight (tracing the inner region of diffuse clouds with weak UV fields). \textcolor{black}{[11] Such environmental dependence of λλ5780.5 and 5797.1 are also
shown by the recent DIB mapping studies in various environments by \citet{van09} for extra-planar gas, \citet{bai15a} for LMC, and \citet{bai15b} and \citet{far15} for the Local Bubble.} By analogy to the $\lambda$5780.5 carrier, which appears to be a cation from the dependence of its EW on the ambient UV radiation field \citep{son97,cam97,vos11}, we suggest that the family of NIR DIBs also arise from cations. This idea is compatible with the DIB-PAH hypothesis, which proposes that the ionized PAHs are the carriers \textcolor{black}{which} have their transitions in NIR wavelength range \citep{rui05,mat05,cox14}. Recently, in the first conclusive identification of DIB carriers, two DIBs at $\lambda =9577, 9632$ \r{A} were assigned to the absorption features of interstellar ionized buckminsterfullerene (C$_{60}^+$) \citep{foi94,cam15}. Ionized fullerenes and their derivatives are also expected as the carrier sources of NIR DIBs. 

\textcolor{black}{In \textsection{5.2}, we suggested that the $\lambda$10504 carrier is destroyed by the ionizing photons in Cyg OB2, whereas those of the family DIBs can survive the Cyg OB2 environment. We also suggested that the $\lambda$12623 carrier may be produced in Cyg OB2 (\textsection{5.2}).}
\textcolor{black}{The destruction of the $\lambda$10504 carrier and possible production of the $\lambda$12623 carrier are probably caused by the strong UV flux in the Cyg OB2 environment. The different environmental dependence of these two DIBs and the family DIBs would be due to the difference of their carrier properties, such as the ionization state and the stability against the UV photon flux.}

\textcolor{black}{The full width at half maximum (FWHM) of $\lambda$10504 is $\sim$ 1.2 cm$^{-1}$, which is one of the smallest widths among 17 DIBs investigated in this paper, while the FWHM of $\lambda$12623 is $\sim$ 2.4 cm$^{-1}$, which is one of the largest. If the profiles of the DIBs are unresolved rotational contours, the widths are mainly determined by the rotational constants of the carriers \citep{cam04,oka13}. Assuming that the geometries of molecules are similar, the band widths of the smaller molecules should be broader due to the larger rotational constants. Therefore, the $\lambda$10504 carrier would be larger than the $\lambda$12623 carrier. Then, we speculate that the photodissociation of the $\lambda$10504 carrier results in the production of $\lambda$12623 carrier. }

\section{Summary}

We investigated the environmental dependence of DIBs toward Cyg OB2 using WINERED spectrograph data. We observed the seven brightest OB stars in the $J$ band in Cyg OB2, and detected all known NIR DIBs with large EWs toward all observed stars in the wavelength coverage of WINERED. Our findings are summarized below:

\begin{enumerate}

\item Five of the DIBs ($\lambda \lambda$10780, 10792, 11797, 12623, and 13175) are well correlated, suggesting similar physical and/or chemical properties of their carriers.

\item The EWs of DIB $\lambda$10504, which were suggested to be well correlated with the family of NIR DIBs in Paper I, are almost constant toward Cyg OB2. In contrast, the EWs of the family of NIR DIBs vary by a factor of two. We attribute the different behaviors of the DIB $\lambda$10504 and the family to the environmental sensitivity of the DIB $\lambda$10504 carrier.

\item None of the NIR DIBs are correlated with the C$_2$ column densities, suggesting that the NIR DIB carriers are distributed mainly in the diffuse component rather than the dense component, in which the C$_2$ molecules are distributed. 

\item Plotting the EWs of DIB $\lambda$10504 \textcolor{black}{and the family DIBs} as a function of $E(B-V)$, we found that $\lambda$10504 toward Cyg OB2 derivates from the linear correlation derived from the data in Paper I, becoming weaker than expected. This suggests that $\lambda$10504 differs from the family NIR DIBs in Cyg OB2 because its carrier is removed. According to the regression line in the plot, the average EW of $\lambda$10504 (78.4$\pm$5.4 m\r{A}) corresponds to $A_V \sim 3.6$mag, similar to the lower limit of the extinction distribution of OB stars in Cyg OB2. Therefore, the $\lambda$10504 carrier seems to be completely destroyed in Cyg OB2. 
Considering the high UV radiation field in Cyg OB2, we surmise that the carrier is removed by processes such as dissociation and ionization.

\end{enumerate}

We are grateful to the staff of Koyama Astronomical Observatory for their support during our observation. \textcolor{black}{We also thank the referee for a careful reading of the manuscript and helpful comments that improved this paper.} This study is financially supported by KAKENHI (16684001) Grant-in-Aid
for Young Scientists (A), KAKENHI (20340042) Grant-in-Aid for Scientific
Research (B), KAKENHI (26287028) Grant-in-Aid for Scientific
Research (B), KAKENHI (21840052) Grant-in-Aid for Young Scientists
(Start-up), and Japan Society for the Promotion of Science, MEXT
(Ministry of Education, Culture, Sports, Science and Technology, Japan)
- Supported program for the Strategic Research Foundation at Private
Universities, 2008 - 2012, S0801061. S.H. was supported by JSPS Fellows Grant Number 25-10504.

\end{document}